\documentclass[prd,superscriptaddress,showpacs,nofootinbib,amsmath,amssymb,aps,11pt]{revtex4}

\usepackage{bm}
\usepackage{amsfonts}
\usepackage{latexsym}
\usepackage[latin1]{inputenc}
\usepackage{graphicx}
\usepackage{amsmath}
\usepackage{palatino}
\usepackage{mathpazo}
\usepackage{booktabs}
\usepackage{dcolumn}

\def\jnl@style{\it}
\def\aaref@jnl#1{{\jnl@style#1}}

\def\aaref@jnl#1{{\jnl@style#1}}

\def\aj{\aaref@jnl{AJ}}                   % Astronomical Journal
\def\apj{\aaref@jnl{ApJ}}                 % Astrophysical Journal
\def\apjl{\aaref@jnl{ApJ}}                % Astrophysical Journal, Letters
\def\apjs{\aaref@jnl{ApJS}}               % Astrophysical Journal, Supplement
\def\apss{\aaref@jnl{Ap\&SS}}             % Astrophysics and Space Science
\def\aap{\aaref@jnl{A\&A}}                % Astronomy and Astrophysics
\def\aapr{\aaref@jnl{A\&A~Rev.}}          % Astronomy and Astrophysics Reviews
\def\aaps{\aaref@jnl{A\&AS}}              % Astronomy and Astrophysics, Supplement
\def\mnras{\aaref@jnl{Mon.~Not.~Roy.~Astron.~Soc.}}             % Monthly Notices of the RAS
\def\prd{\aaref@jnl{Phys.~Rev.~D}}        % Physical Review D
\def\prc{\aaref@jnl{Phys.~Rev.~C}}  % Physical Review C
\def\prl{\aaref@jnl{Phys.~Rev.~Lett.}}    % Physical Review Letters
\def\qjras{\aaref@jnl{QJRAS}}             % Quarterly Journal of the RAS
\def\skytel{\aaref@jnl{S\&T}}             % Sky and Telescope
\def\ssr{\aaref@jnl{Space~Sci.~Rev.}}     % Space Science Reviews
\def\zap{\aaref@jnl{ZAp}}                 % Zeitschrift fuer Astrophysik
\def\nat{\aaref@jnl{Nature}}              % Nature
\def\aplett{\aaref@jnl{Astrophys.~Lett.}} % Astrophysics Letters
\def\apspr{\aaref@jnl{Astrophys.~Space~Phys.~Res.}} % Astrophysics Space Physics Research
\def\physrep{\aaref@jnl{Phys.~Rep.}}      % Physics Reports
\def\physscr{\aaref@jnl{Phys.~Scr}}       % Physica Scripta
\def\commat{\aaref@jnl{Comm.~Math.~Phys.}}              % Communications in Mathematical Physics
\def\science{\aaref@jnl{Science}}               % Science
\def\cqg{\aaref@jnl{Classical Quant.~Grav.}}            % Classical and Quantum Gravity
\def\jpcs{\aaref@jnl{JPCS}}                                     % Journal of Physics Conference Series
\def\ijmpd{\aaref@jnl{Int.~J.~Mod.~Phys.~D}}                    % International Journal of Modern Physics D
\def\grg{\aaref@jnl{Gen.~Relat.~Gravit.}}               % General Relativity and Gravitation
\def\rpp{\aaref@jnl{Rep.~Prog.~Phys.}}          % Reports on Progress in Physics
\def\npa{\aaref@jnl{Nucl.~Phys.~A}}        % Nuclear Physics A
\def\lrr{\aaref@jnl{Living Rev.~Rel.}}                   % Living reviews in relativity
\def\jcap{\aaref@jnl{J.~Cosmology Astropart.~Phys.}}    % Journal of cosmology and astroparticle physics
\def\rmp{\aaref@jnl{Rev.~Mod.~Phys.}}   %Reviews of modern physics

\allowdisplaybreaks[1]
\begin{document}

\title{New Gauss-Bonnet black holes with curvature induced scalarization in the extended scalar-tensor theories }

\author{Daniela D. Doneva}
\email{daniela.doneva@uni-tuebingen.de}
\affiliation{Theoretical Astrophysics, Eberhard Karls University of T\"ubingen, T\"ubingen 72076, Germany}
\affiliation{INRNE - Bulgarian Academy of Sciences, 1784  Sofia, Bulgaria}

\author{Stoytcho S. Yazadjiev}
\email{yazad@phys.uni-sofia.bg}
\affiliation{Theoretical Astrophysics, Eberhard Karls University of T\"ubingen, T\"ubingen 72076, Germany}
\affiliation{Department of Theoretical Physics, Faculty of Physics, Sofia University, Sofia 1164, Bulgaria}

%%%%%%%%%%%%%%%%%%%%%%%%%%%%%%%%%%%%  DATE  %%%%%%%%%%%%%%%%%%%%%%%%%%%%%%%%%%%%
%\date{\today}

\begin{abstract}
In the present paper we consider a class of extended scalar-tensor-Gauss-Bonnet (ESTGB) theories  for which the scalar degree of freedom is excited only in the extreme curvature regime.  We  show that in the mentioned class of ESTGB theories there exist  new black hole solutions which are formed by spontaneous scalarization of the Schwarzaschild balck holes in the extreme curvature regime. 
In this regime, below certain mass, the Schwarzschild solution becomes unstable and new branch of solutions with nontrivial scalar field bifurcate from the Schwarzschild one. As a matter of fact, more than one branches with nontrivial scalar field can bifurcate at different masses but only the first one is supposed to be stable. This effect is quite similar to the spontaneous scalarization of neutron stars. In contrast with the standard spontaneous scalarization of neutron stars  which is induced by the presence of matter, in our case the scalarization is induced by the curvature of the spacetime.	
\end{abstract}


\maketitle

\section{Introduction}

The historic  direct detection of gravitational waves has opened a new era in physics, giving a powerful tool for  exploring  the strong-gravity regime, where spacetime curvature is extreme. General Relativity is well-tested in the weak-field regime, whereas the strong-field regime still remains essentially unexplored and unconstrained. There are both phenomenological and theoretical reasons for the modification of the original Einstein  quations. For example, predictions based on General Relativity and the Standard Model of particle physics fail to explain the accelerated expansion of the Universe.  
It is also well known that the pure general relativity is not a renormalizable theory which poses severe obstacles to the efforts  of quantizing gravity. 
The renormalization at one loop demands that the Einstein-Hilbert action be supplemented with all the possible algebraic curvature invariants of second order \cite{Stelle_1977}. On the other hand, the attempts to construct a unified theory of all the interactions, naturally lead to scalar-tensor type generalizations of general relativity with an additional dynamical scalar field and with Lagrangians containing various kinds of curvature corrections to the usual Einstein-Hilbert Lagrangian coupled to the scalar field \cite{Berti_2015}-\cite{Pani_2011a}. The most natural modifications of this class are the extended scalar-tensor theories (ESTT) where the usual Einstein-Hilbert action is supplemented with all possible algebraic curvature invariants of second order with a dynamical scalar field nonminimally coupled to these invariants

The equations of the ESTT in their most general form are of order higher than two. This in general can lead to the Ostrogradski instability and to the appearance of ghosts. However, there is a particular sector of the ESTT, namely the sector where the scalar field is coupled exactly to the Gauss-Bonnet invariant, for which the field equations are of second order as in general relativity and the theory is free from ghosts. Due to these reasons, in the present paper we shall focus on the extended scalar-tensor-Gauss-Bonnet (ESTGB) gravity as a
natural modification of general relativity and a natural extension of the standard scalar-tensor theories. 

A particular model of ESTGB gravity, the so-called Einstein-dilaton-Gauss-Bonnet(EdGB) gravity, was extensively studied in the literature.
The nonrotating  black holes in EdGB gravity with a coupling function
$\alpha e^{\gamma \varphi}$ and vanishing potential for the dilaton field, with  $\alpha$ and $\gamma$ being constants, were studied perturbatively or numerically in \cite{Mignemi_1993}-\cite{Pani_2009}. It was shown that the EdGB black holes exist only for $\alpha>0$ and when the black hole mass is greater than certain lower bound proportional to the paremeter $\alpha$. The slowly rotating EdGB black holes were studied in \cite{Pani_2009}, \cite{Ayzenberg_2014} and \cite{Maselli_2015}. The rapidly rotating EDGB black holes were constructed
numerically in \cite{Kleihaus_2011}-\cite{Kleihaus_2016}. The rotating EdGB black holes can exist only when the mass and the angular momentum fall in certain domain depending on the coupling constant. Another interesting fact 
about the EdGB black holes is that they can exceed the Kerr bound for the angular momentum.

In the present paper we shall consider 
a class of ESTGB theories with a scalar coupling functions for which the scalar degree of freedom is excited only in the extreme curvature regime.  In particular we shall show that in the mentioned class of ESTGB theories there exist  new black hole solutions 
which are formed by spontaneous scalarization of the Schwarzaschild balck holes in the extreme curvature regime. In contrast with 
the standard spontaneous scalarization \cite{Damour_1993}-\cite{Doneva_2010} which is induced by the presence of matter, in our case the scalarization is induced by the curvature of the spacetime.

\section{Basic equations and setting the problem}

The general action of ESTGB theories in vacuum  is given by  

\begin{eqnarray}
S=&&\frac{1}{16\pi}\int d^4x \sqrt{-g} 
\Big[R - 2\nabla_\mu \varphi \nabla^\mu \varphi - V(\varphi) 
 + \lambda^2 f(\varphi){\cal R}^2_{GB} \Big] ,\label{eq:quadratic}
\end{eqnarray}
where $R$ is the Ricci scalar with respect to the spacetime metric $g_{\mu\nu}$, $\varphi$ is the scalar field with a potential $V(\varphi)$ and a coupling function  $f(\varphi)$ depending only on $\varphi$, $\lambda$ is the Gauss-Bonnet coupling constant having  dimension of $length$ and ${\cal R}^2_{GB}$ is the Gauss-Bonnet invariant\footnote{The Gauss-Bonnet invariant is defined by ${\cal R}^2_{GB}=R^2 - 4 R_{\mu\nu} R^{\mu\nu} + R_{\mu\nu\alpha\beta}R^{\mu\nu\alpha\beta}$ where $R$ is the Ricci scalar, $R_{\mu\nu}$ is the Ricci tensor and $R_{\mu\nu\alpha\beta}$ is the Riemann tensor}. The action yields the following field equations

\begin{eqnarray}\label{FE}
&&R_{\mu\nu}- \frac{1}{2}R g_{\mu\nu} + \Gamma_{\mu\nu}= 2\nabla_\mu\varphi\nabla_\nu\varphi -  g_{\mu\nu} \nabla_\alpha\varphi \nabla^\alpha\varphi - \frac{1}{2} g_{\mu\nu}V(\varphi),\\
&&\nabla_\alpha\nabla^\alpha\varphi= \frac{1}{4} \frac{dV(\varphi)}{d\varphi} -  \frac{\lambda^2}{4} \frac{df(\varphi)}{d\varphi} {\cal R}^2_{GB},
\end{eqnarray}
where  $\nabla_{\mu}$ is the covariant derivative with respect to the spacetime metric $g_{\mu\nu}$ and  $\Gamma_{\mu\nu}$ is defined by 

\begin{eqnarray}
\Gamma_{\mu\nu}&=& - R(\nabla_\mu\Psi_{\nu} + \nabla_\nu\Psi_{\mu} ) - 4\nabla^\alpha\Psi_{\alpha}\left(R_{\mu\nu} - \frac{1}{2}R g_{\mu\nu}\right) + 
4R_{\mu\alpha}\nabla^\alpha\Psi_{\nu} + 4R_{\nu\alpha}\nabla^\alpha\Psi_{\mu} \nonumber \\ 
&& - 4 g_{\mu\nu} R^{\alpha\beta}\nabla_\alpha\Psi_{\beta} 
 + \,  4 R^{\beta}_{\;\mu\alpha\nu}\nabla^\alpha\Psi_{\beta} 
\end{eqnarray}  
with 

\begin{eqnarray}
\Psi_{\mu}= \lambda^2 \frac{df(\varphi)}{d\varphi}\nabla_\mu\varphi .
\end{eqnarray}

In what follows we shall focus on the case $V(\varphi)=0$.

We consider further static and spherically symmetric spacetimes as well as static and spherically symmetric 
scalar field configurations. The spacetime metric then can be written in the standard form

\begin{eqnarray}
ds^2= - e^{2\Phi(r)}dt^2 + e^{2\Lambda(r)} dr^2 + r^2 (d\theta^2 + \sin^2\theta d\phi^2 ). 
\end{eqnarray}   

The dimensionally reduced field equations (\ref{FE}) are the following

\begin{eqnarray}
&&\frac{2}{r}\left[1 +  \frac{2}{r} (1-3e^{-2\Lambda})  \Psi_{r}  \right]  \frac{d\Lambda}{dr} + \frac{(e^{2\Lambda}-1)}{r^2} 
- \frac{4}{r^2}(1-e^{-2\Lambda}) \frac{d\Psi_{r}}{dr} - \left( \frac{d\varphi}{dr}\right)^2=0, \label{DRFE1}\\ && \nonumber \\
&&\frac{2}{r}\left[1 +  \frac{2}{r} (1-3e^{-2\Lambda})  \Psi_{r}  \right]  \frac{d\Phi}{dr} - \frac{(e^{2\Lambda}-1)}{r^2} - \left( \frac{d\varphi}{dr}\right)^2=0,\label{DRFE2}\\ && \nonumber \\
&& \frac{d^2\Phi}{dr^2} + \left(\frac{d\Phi}{dr} + \frac{1}{r}\right)\left(\frac{d\Phi}{dr} - \frac{d\Lambda}{dr}\right)  + \frac{4e^{-2\Lambda}}{r}\left[3\frac{d\Phi}{dr}\frac{d\Lambda}{dr} - \frac{d^2\Phi}{dr^2} - \left(\frac{d\Phi}{dr}\right)^2 \right]\Psi_{r} 
\nonumber \\ 
&& \hspace{0.5cm} - \frac{4e^{-2\Lambda}}{r}\frac{d\Phi}{dr} \frac{d\Psi_r}{dr} + \left(\frac{d\varphi}{dr}\right)^2=0, \label{DRFE3}\\ && \nonumber \\
&& \frac{d^2\varphi}{dr^2}  + \left(\frac{d\Phi}{dr} \nonumber - \frac{d\Lambda}{dr} + \frac{2}{r}\right)\frac{d\varphi}{dr} \nonumber \\ 
&& \hspace{0.5cm} - \frac{2\lambda^2}{r^2} \frac{df(\varphi)}{d\phi}\Big\{(1-e^{-2\Lambda})\left[\frac{d^2\Phi}{dr^2} + \frac{d\Phi}{dr} \left(\frac{d\Phi}{dr} - \frac{d\Lambda}{dr}\right)\right]    + 2e^{-2\Lambda}\frac{d\Phi}{dr} \frac{d\Lambda}{dr}\Big\} =0, \label{DRFE4}\\ 
\end{eqnarray}
with 

\begin{eqnarray}
\Psi_{r}=\lambda^2 \frac{df(\varphi)}{d\varphi} \frac{d\varphi}{dr}.
\end{eqnarray}
 
In the present paper we are interested in ESTGBT with coupling function $f(\varphi)$ satisfying the conditions\footnote{We consider here the case when the cosmological value of the scalar field is zero, $ \varphi_{\infty}=0$. } 
$\frac{df}{d\varphi}(0)=0$ and $b^2=\frac{d^2f}{d\varphi^2}(0)>0$.  Without loss of generality we can  put $b=1$  and this can be achieved by rescaling  the coupling parameter $\lambda\to b\lambda$ and by redefining  the coupling function $f\to b^{-2}f$. In addition, since the theory depends only on $\frac{df(\varphi)}{d\varphi}$, we can also impose $f(0)=0$. 

The natural and the important question is whether the class of ESTGBT defined above admits (static and spherically symmetric) black hole solutions.  From the dimensionally reduced field equations \eqref{DRFE1}--\eqref{DRFE4} it is clear that the usual Schwarzschlild black hole solution is also a black hole solution to the ESTGBT under consideration with a trivial scalar field $\varphi=0$.  We  shall however show that the Schwarzschild solution within the certain range of the mass  is unstable in the framework of the ESTGBT under consideration. For this purpose we consider the perturbations of the Schwarzschild solution with mass $M$ within the framework of the described class of ESTGBT. It is not difficult to see that in the considered class of ESTGBT the equations governing the perturbations of the metric $\delta g_{\mu\nu}$ are decoupled from the equation governing the perturbation $\delta \varphi$ of the scalar field. The equations for metric perturbations  are in fact the same as those in the pure Einstein gravity and therefore we shall focus only on the scalar field perturbations. The equation governing the 
scalar perturbations is    
 
\begin{eqnarray}\label{PESF}
\Box_{(0)} \delta\varphi + \frac{1}{4}\lambda^2  {\cal R}^2_{GB(0)} \delta\varphi=0, 
\end{eqnarray} 
where $\Box_{(0)}$ and ${\cal R}^2_{GB(0)}$ are the D'alambert operator and the Gauss-Bonnet invariant for 
the Schwarzschild geometry. Taking into account that the  background geometry  is static and spherically
symmetric, the variables can be separated in the following way

\begin{eqnarray}
\delta\varphi= \frac{u(r)}{r} e^{-i\omega t}Y_{lm}(\theta,\phi),
\end{eqnarray}
with $Y_{lm}(\theta,\phi)$ being the spherical harmonics.  After substituting
in (\ref{PESF}) we find

\begin{eqnarray}
\frac{g(r)}{r} \frac{d}{dr}\left[ r^2 g(r) \frac{d}{dr}\left(\frac{u(r)}{r}\right)\right] 
+ \left[\omega^2 + g(r)\left( -\frac{l(l+1)}{r^2} + \frac{1}{4}\lambda^2  {\cal R}^2_{GB(0)}\right)\right]u(r)=0,
\end{eqnarray}
where $g(r)=1-\frac{2M}{r}$ and ${\cal R}^2_{GB(0)}= \frac{48M^2}{r^6}$ for the Schwarzschild solution. By introducing the tortoise coordinate
$dr_{*}=\frac{dr}{g(r)}$ which maps the domain $r\in (2M,+\infty)$ to $r_{*}\in (-\infty,+\infty)$, the equation can be cast in the Schr\"odinger form

 \begin{eqnarray}
\frac{d^2u}{dr^2_{*}} + [ \omega^2 - U(r)]u=0 \label{eq:PerturbEq}
 \end{eqnarray}
  	
with a potential 
 \begin{eqnarray}
U(r)= \left(1-\frac{2M}{r}\right)\left[\frac{2M}{r^3} + \frac{l(l+1)}{r^2} - \lambda^2 \frac{12M^2}{r^6}\right].  
\end{eqnarray}

A sufficient condition for the existence of an unstable mode is \cite{Buell_1995} 

\begin{eqnarray}
\int_{-\infty}^{+\infty} U(r_{*}) dr_{*}=\int^{\infty}_{2M} \frac{U(r)}{1-\frac{2M}{r}}dr <0.   
\end{eqnarray}  

For the spherically symmetric perturbations the above condition gives $M^2<\frac{3}{10}\lambda^2 $. Therefore we can conclude that the Schwarzshild black holes
with mass satisfying  $M^2<\frac{3}{10}\lambda^2 $ are unstable within the framework of the ESTGBT under consideration. Stated differently, 
the Schwarzshild black holes become unstable when the curvature of the horizon exceeds  a certain critical value -- in terms of the Kretschmann scalar of the horizon ${\cal K}_{H}$, the instability occurs when ${\cal K}_{H}> \frac{25}{3\lambda^4}$.

This result naturally leads us to the conjecture that, in our class of ESTGBT and in the interval where the Schwarzschild solution is unstable,  there exist black solutions with nontrivial scalar field.  In the next sections we numerically prove that such black hole solutions really exist and present some of their basic properties.   

\section{Numerical setup}
In order to obtain the black hole solutions with a nontrivial scalar field we solve numerically the system of reduced field equations \eqref{DRFE1}--\eqref{DRFE4}. The system of differential equations is transformed in the following way in order to simplify the numerical calculations. Using equations \eqref{DRFE1} and \eqref{DRFE2} one can straightforward obtain expressions for the metric function $\Lambda$ and its derivative that depend only on the metric function $\Phi$, the scalar field $\varphi$ and their derivatives. In this way the system of equations \eqref{DRFE3}--\eqref{DRFE4} decouples from the rest of the equations and we are left with two second order partial differential equations for $\Phi$ and $\varphi$. Let us point out that after a solution for $\Phi$ and $\varphi$ is found the metric function $\Lambda$ can be constructed directly without the need for integration due to the particular form of equation \eqref{DRFE2}. 

One can notice that in all of the reduced field equations the metric function $\Phi$ do not enter directly but only through its derivatives. That is why instead of solving two second order equations one can solve one second order equation for the scalar field $\varphi$ and one first order equation for the first derivative of the metric function $d\Phi/dr$. Later $\Phi$ can be found by simply integrating the resulting  $d\Phi/dr$ with the appropriate boundary conditions. Even though this might look like a very small simplification it is very important since it reduces the number of the shooting parameters from two to one and thus simplifies a lot the search for bifurcations of the Schwarzschild solution.

Let us discuss now the boundary and the regularity conditions. They come from the requirements for asymptotic flatness at infinity and 
the regularity at the black hole horizon $r=r_H$. 
As usual the asymptotic flatness imposes the following asymptotic conditions  

\begin{eqnarray}
\Phi|_{r\rightarrow\infty} \rightarrow 0, \;\;  \Lambda|_{r\rightarrow\infty} \rightarrow 0,\;\; \varphi|_{r\rightarrow\infty} \rightarrow 0\;\;.   \label{eq:BH_inf}
\end{eqnarray} 
The very existence of black hole horizon requires 

\begin{eqnarray}
e^{2\Phi}|_{r\rightarrow r_H} \rightarrow 0, \;\; e^{-2\Lambda}|_{r\rightarrow r_H} \rightarrow 0. \label{eq:BC_rh}
\end{eqnarray} 

The regularity of the scalar field and its first  and second derivatives on the black hole horizon
gives one more condition, namely 

\begin{eqnarray}\label{QE}
\left(\frac{d\varphi}{dr}\right)_{H} + \frac{2\lambda^2}{r_{H}} \frac{df}{d\varphi}(\varphi_{H}) \left(\frac{d\varphi}{dr}\right)^2_{H} + 
\frac{2\lambda^2}{r^3_{H}} \frac{df}{d\varphi}(\varphi_{H})=0.
\end{eqnarray} 

From  (\ref{QE}) we can express the value of the first derivative $\frac{df}{d\varphi}(\varphi_{H})$ as a function of the value of scalar field on the horizon and the horizon radius, namely  
\begin{eqnarray}
\left(\frac{d\varphi}{dr}\right)_{H}= \frac{r_{H}}{4 \lambda^2 \frac{df}{d\varphi}(\varphi_{H})} 
\left[-1 \pm \sqrt{1 - \frac{24\lambda^4}{r^4_{H}} \left(\frac{df}{d\varphi}(\varphi_{H})\right)^2}\right].
\end{eqnarray} 
In this expression we have to  chose the plus sign since only in this case we can recover the trivial solution in the limit $\varphi_{H}\to 0$. The requirement for positiveness of the expression inside the square root in the above expression imposes restriction on the possible solutions with nontrivial scalar field, i.e. black hole solutions exist only when
\begin{equation}
r_H^4 > 24 \lambda^4 \left(\frac{df}{d\varphi}(\varphi_{H})\right)^2. \label{eq:BC_sqrt_rh}
\end{equation}
For Schwarzschild $\varphi=0$ and since the coupling function by definition satisfies $\frac{df}{d\varphi}(0)=0$, this condition is automatically satisfied. For black holes with nontrivial scalar field, though, eq. \eqref{eq:BC_sqrt_rh} can be violated as we will see below.

We use a shooting method to find the black hole solutions. First, the first order differential equation for $d\Phi/dr$ and the second order equation for $\varphi$ are solved using the above boundary conditions. The input parameter which determines the black hole solutions is $r_H$ and we have one shooting parameter -- the value of the scalar field at the horizon. This shooting parameter is determined by the boundary condition of $\varphi$ at infinity \eqref{eq:BH_inf}.  The metric function $\Phi$ is calculated afterwards also with a shooting procedure using the obtained $d\Phi/dr$ and the boundary conditions \eqref{eq:BH_inf}. As we commented, after we have found the solutions for $\Phi$ and $\varphi$, $\Lambda$ can be determined directly without the need for integration using eq. \eqref{DRFE2}. The mass of the black hole $M$ and the dilaton charge $D$ are obtained through the asymptotics of the functions $\Lambda$, $\Phi$ and $\varphi$, namely 
\begin{equation}
\Lambda\approx \frac{M}{r} + O(1/r^2), \;\; \Phi\approx  -\frac{M}{r} + O(1/r^2), \; \; \varphi\approx \frac{D}{r} + O(1/r^2).
\end{equation}

\section{Results}
The only thing that remains to be fixed is the explicit form of the coupling function $f(\varphi)$. In the present paper 
we consider the following function 
\begin{equation}
f(\varphi)=  \frac{1}{12} \left[1- \exp(-6\varphi^2)\right], \label{eq:coupling_function}
\end{equation} 
that is chosen in such a way that we have both non-negligible deviations from the Schwarzschild solution and the condition for the existence of solutions with nontrivial scalar field \eqref{eq:BC_sqrt_rh} is fulfilled for large enough range of parameters. We have explicitly checked that other choices of $f(\varphi)$ that lead to similar results are of course possible but exploring a large variety of $f(\varphi)$ functions is out of the scope of the present paper. Instead, we will focus on examining in detail the black hole solutions with non-trivial scalar field and the non-uniqueness of the solutions.

As discussed above, the Schwarzschild solution with zero scalar field is always a solution of the field equations but in a certain region of the parameter space it becomes unstable in the framework of the ESTGBT under consideration and new solutions with nontrivial scalar field appear. Moreover, there can be regions where more than one solution with nontrivial scalar field exists and this corresponds roughly speaking to the appearance of more than one bound state of the potential in the perturbation equation\footnote{In the present paper we consider only spherically symmetric solutions and therefore $l=0$.} \eqref{eq:PerturbEq}. The different branches of solutions will have scalar field with different number of zeros similar to the eigenfunctions of the perturbation equation \eqref{eq:PerturbEq}.

Finding the solutions with nontrivial scalar field might be sometimes numerically difficult and it is of great help to know the exact points of bifurcation. In the previous sections we discussed that for $M^2<\frac{3}{10}\lambda^2$ the Schwarzschild black holes are unstable but this is only a sufficient condition for instability and the true point of the first bifurcations is actually at a little bit larger masses. In order to find the points of bifurcation we can use the fact that they are the same as the points where new unstable modes of eq.  \eqref{eq:PerturbEq} appear (for a detailed discussion see \cite{Doneva_2010}). That is why instead of solving the reduced field equations, we can determine the bifurcations points using the perturbation equation \eqref{eq:PerturbEq}, that is numerically easier. Since we are interested in unstable modes, $\omega^2$ should be negative which leads to the fact that the boundary conditions are zero both at the black hole horizon and infinity (for more details and derivation see \cite{Doneva_2010}). Therefore, we have a self-adjoint Sturm-Liouville problem. 

We employed a shooting procedure to find the eigenvalues and the eigenfunction of eq.  \eqref{eq:PerturbEq} and determined the regions of the parameter space where the Schwarzschild solution is stable, where it is unstable and only one unstable mode is present, where two unstable modes are present and so on. This means that we have determined the points of bifurcation of the Schwarschild solution which significantly simplifies the search for black holes with nontrivial scalar field.

\begin{figure}[htb]
	\includegraphics[width=0.495\textwidth]{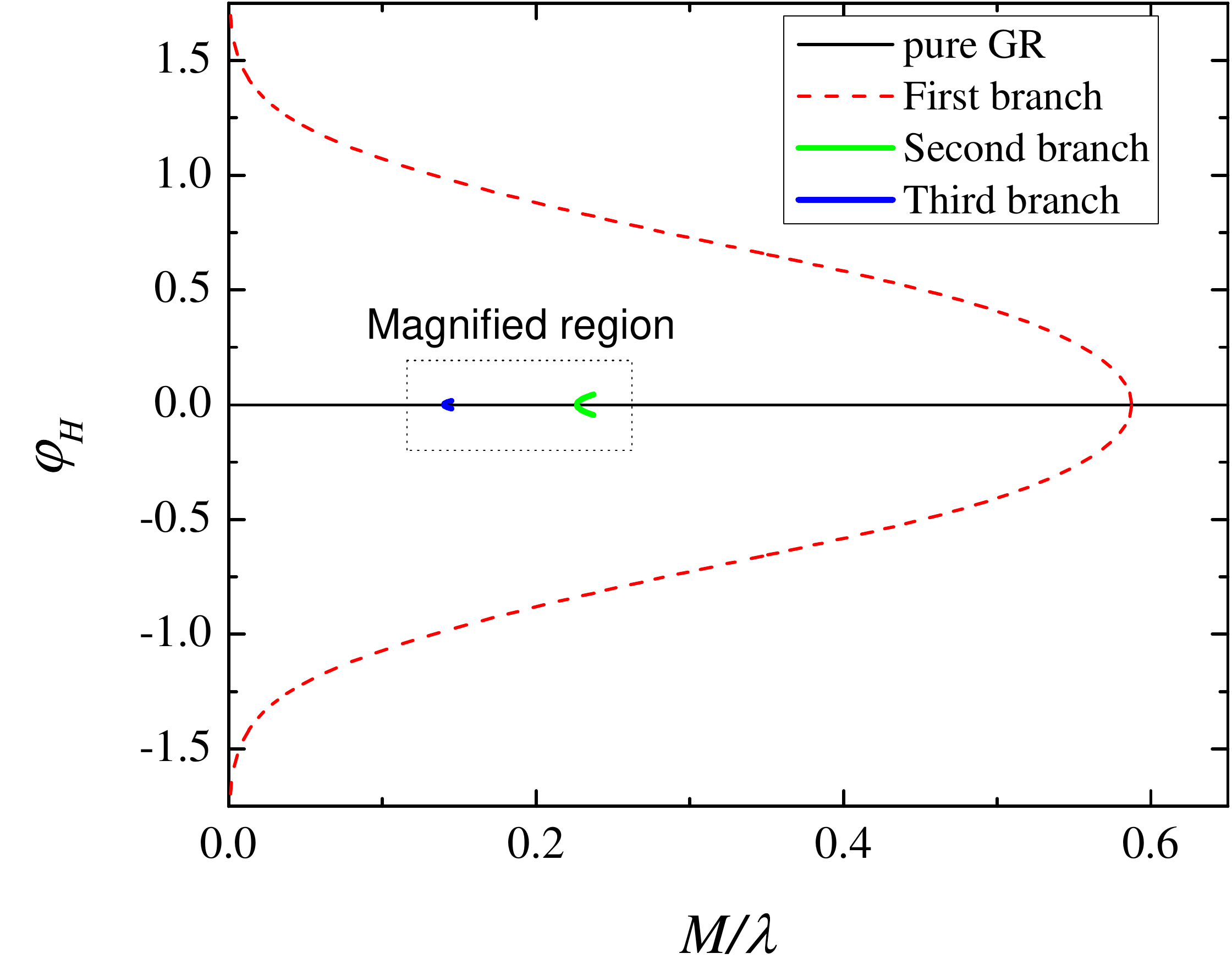}
	\includegraphics[width=0.495\textwidth]{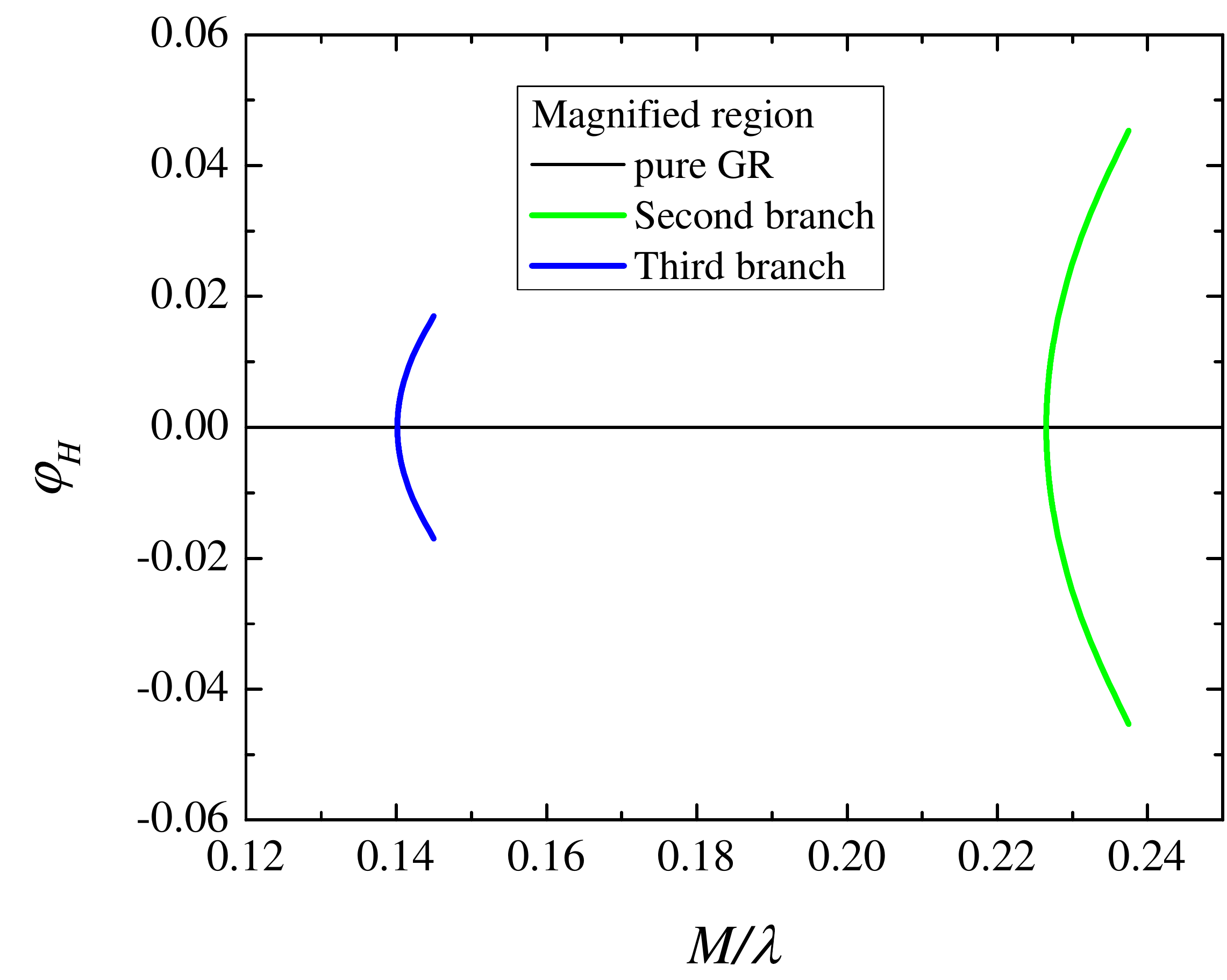}
	\caption{The scalar field at the horizon as a function of the black hole mass. The right figure is a magnification of the left one.}
	\label{fig:phiH(M)}
\end{figure}

\begin{figure}[htb]
	\includegraphics[width=0.495\textwidth]{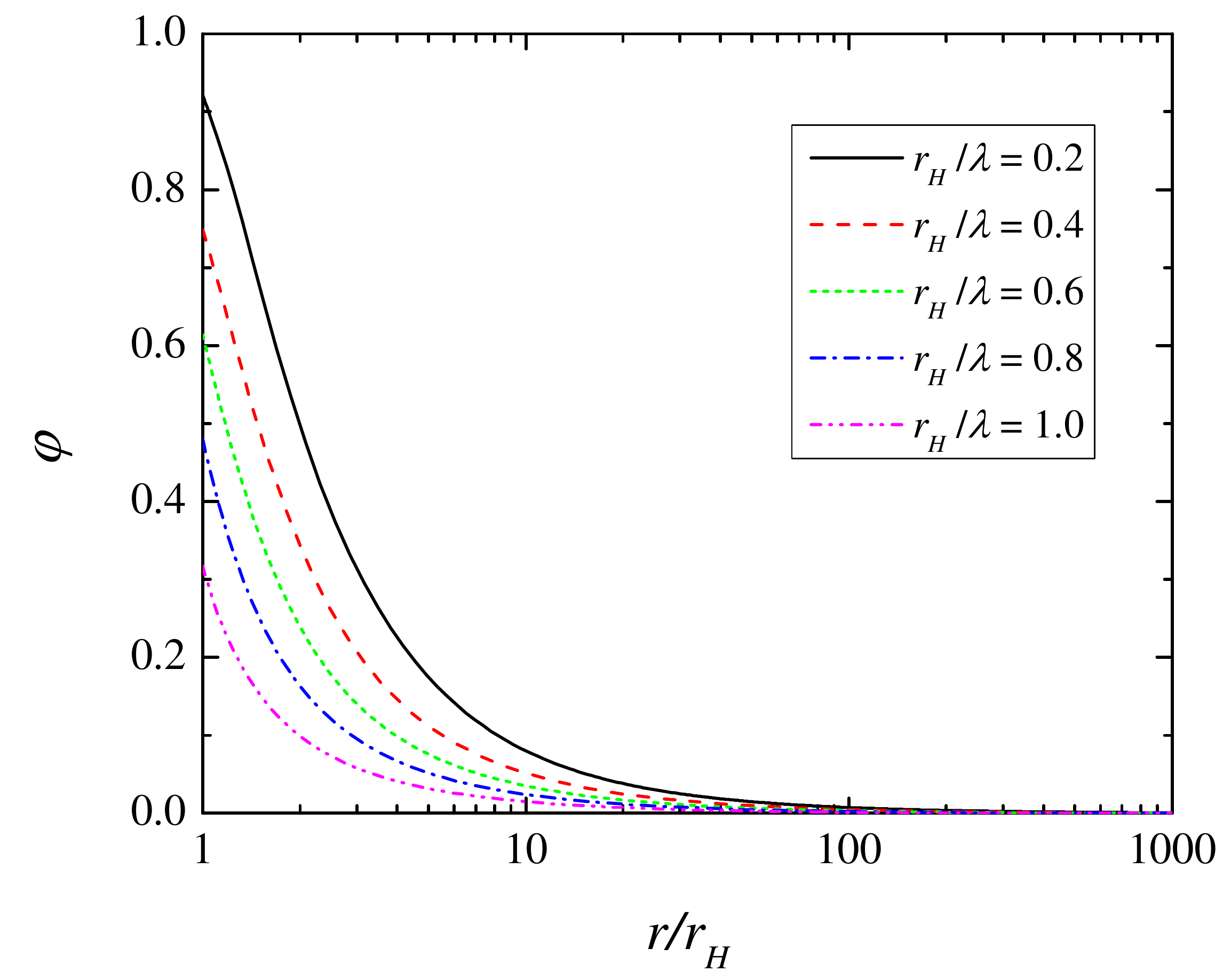}
	\includegraphics[width=0.495\textwidth]{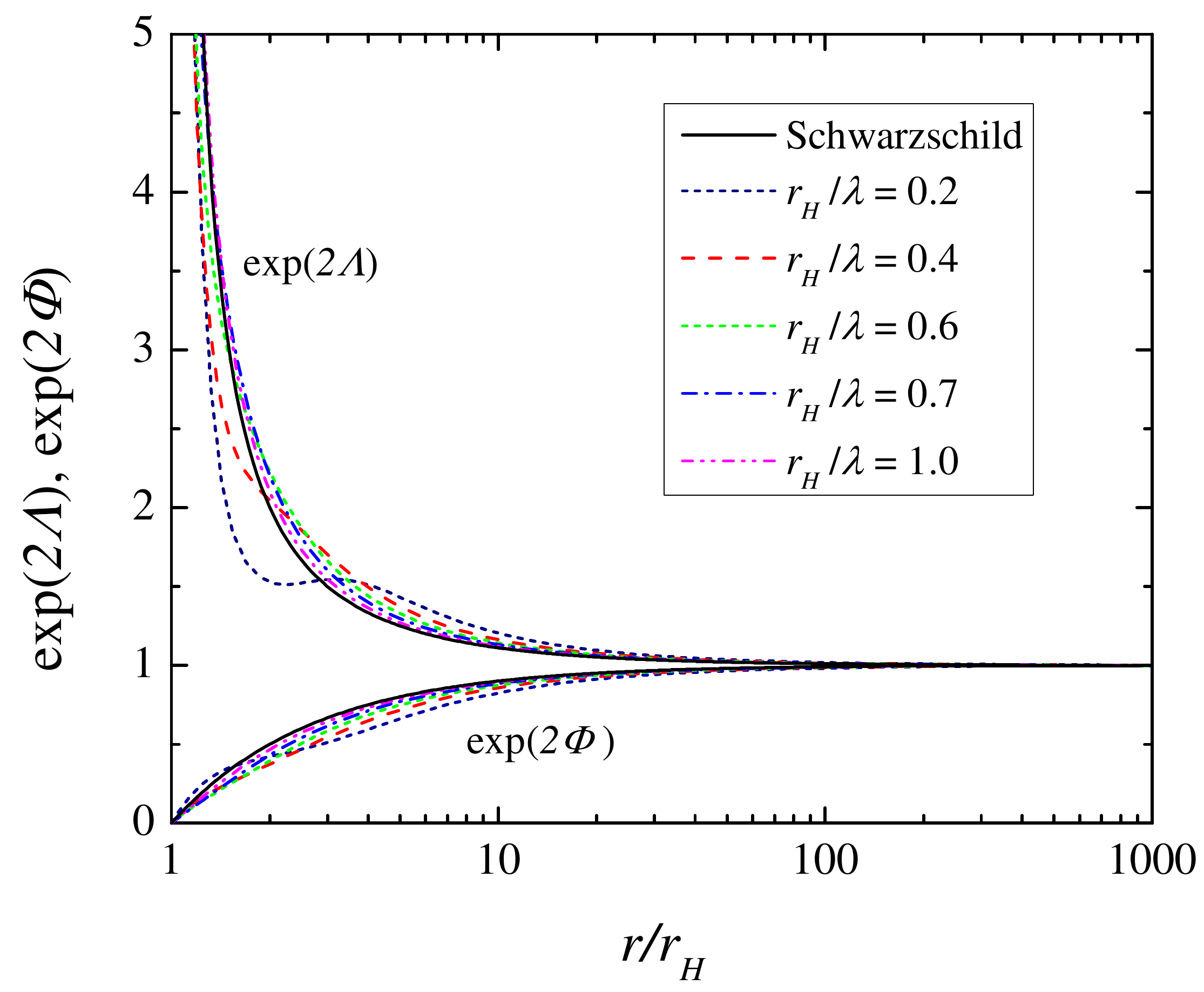}
	\caption{The scalar field and the $g_{tt}$ and $g_{rr}$ components of the metric as  functions of the normalized radial coordinate $r/r_H$ for several black hole solutions that belong to the first nontrivial branch with different values of $r_H/\lambda$.}
	\label{fig:func(r)}
\end{figure}

\begin{figure}[htb]
	\includegraphics[width=0.495\textwidth]{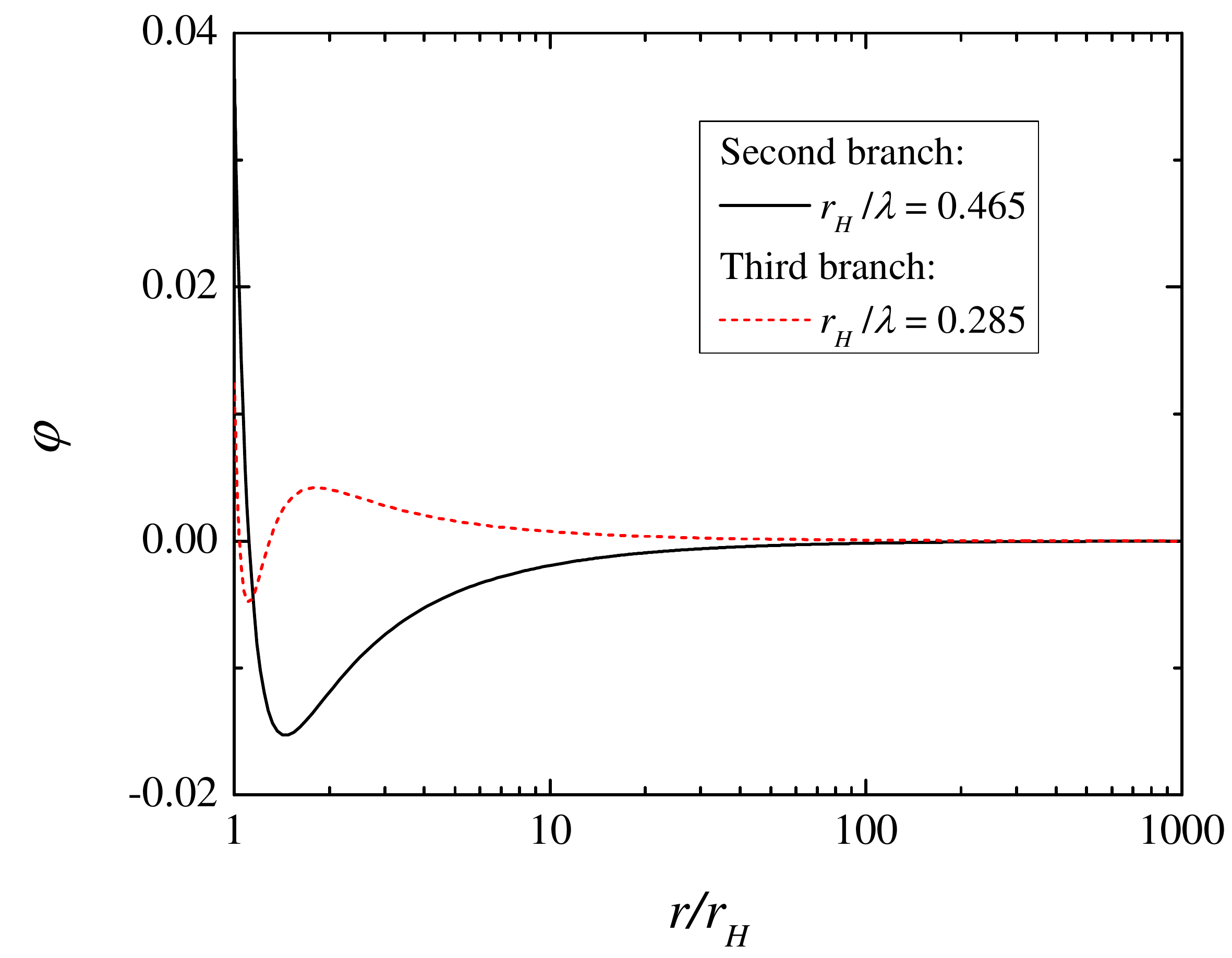}
	\caption{The scalar field as a function of the normalized radial coordinate $r/r_H$ for two representative solution from the second and the third branch of nontrivial solutions. The components of the metric $g_{tt}$ and $g_{rr}$ are not shown since they are almost indistinguishable from the Schwarzschild case.}
	\label{fig:func(r)_Overtones}
\end{figure}

The obtained black hole solutions are plotted in Fig. \ref{fig:phiH(M)} where only the first three bifurcations of the Schwarzschild solution are shown in order to have better visibility.  We will call the Schwarzschild solution the trivial branch of solutions (with trivial scalar field) while the rest of the branches of black holes with nontrivial scalar field will be called nontrivial branches (with nontrivial scalar field). As one can see, all the nontrivial branches start from a bifurcation point at the trivial branch and they span either to $M=0$ (the first nontrivial branch) or they are terminated at some nonzero $M$ (all the other nontrivial branches). The reason for termination of the branches at nonzero $M$ is that beyond this point the condition \eqref{eq:BC_sqrt_rh} is violated. We should point out as well, that calculating black holes with nontrivial scalar field for very small $M$ is very difficult from a numerical points view since the scalar field increases significantly and goes to infinity as $M$ approaches zero. As discussed above, the different nontrivial branches of solutions are characterized by different number of zeros of the scalar field. For the first branch (the red dashed line in Fig. \ref{fig:phiH(M)}) there are no zeros of $\varphi$ as one can see in the left panel of Fig. \ref{fig:func(r)}, the next one (green line) has one zero while the third one (blue line) has two zeros as one can see in Fig. \ref{fig:func(r)_Overtones}. For smaller values of $M$ there are more bifurcation points but our investigations show the corresponding nontivial branches would be even shorter and that is why we have not plotted them. Moreover, it is expected that only the first nontrivial branch characterized by a scalar field without zeros will be stable while the rest of the branches correspond to unstable solutions.

One can notice as well that Fig. \ref{fig:phiH(M)} is symmetric with respect to the x-axis that can be shown easily analytically also using the field equations with the particular coupling function \eqref{eq:coupling_function}. Thus for a fixed $M$ the solutions with positive and negative values of $\varphi_H$ would naturally have opposite signs of the dilaton charge, but they  have the same metric functions and thus mass. The components of the metric $g_{tt}$ and $g_{rr}$, as well as the scalar field  as function of the normalized radial coordinate $r_H/\lambda$ are shown for some representative solutions with different $r_H/\lambda$ in Figs. \ref{fig:func(r)} and \ref{fig:func(r)_Overtones} for the first tree nontrivial branches. These figures demonstrate what we have commented above -- the different branches of nontrivial solutions are characterized by different numbers of zeros of the scalar field. As one can see, the metric of the first nontrivial branch can deviate significantly not only qualitatively but also quantitatively from the Schwarzschild one. We have not plotted $g_{tt}$ and $g_{rr}$ for the next nontrivial branches since they are almost indistinguishable from the pure general relativistic case.  

\begin{figure}[htb]
	\includegraphics[width=0.495\textwidth]{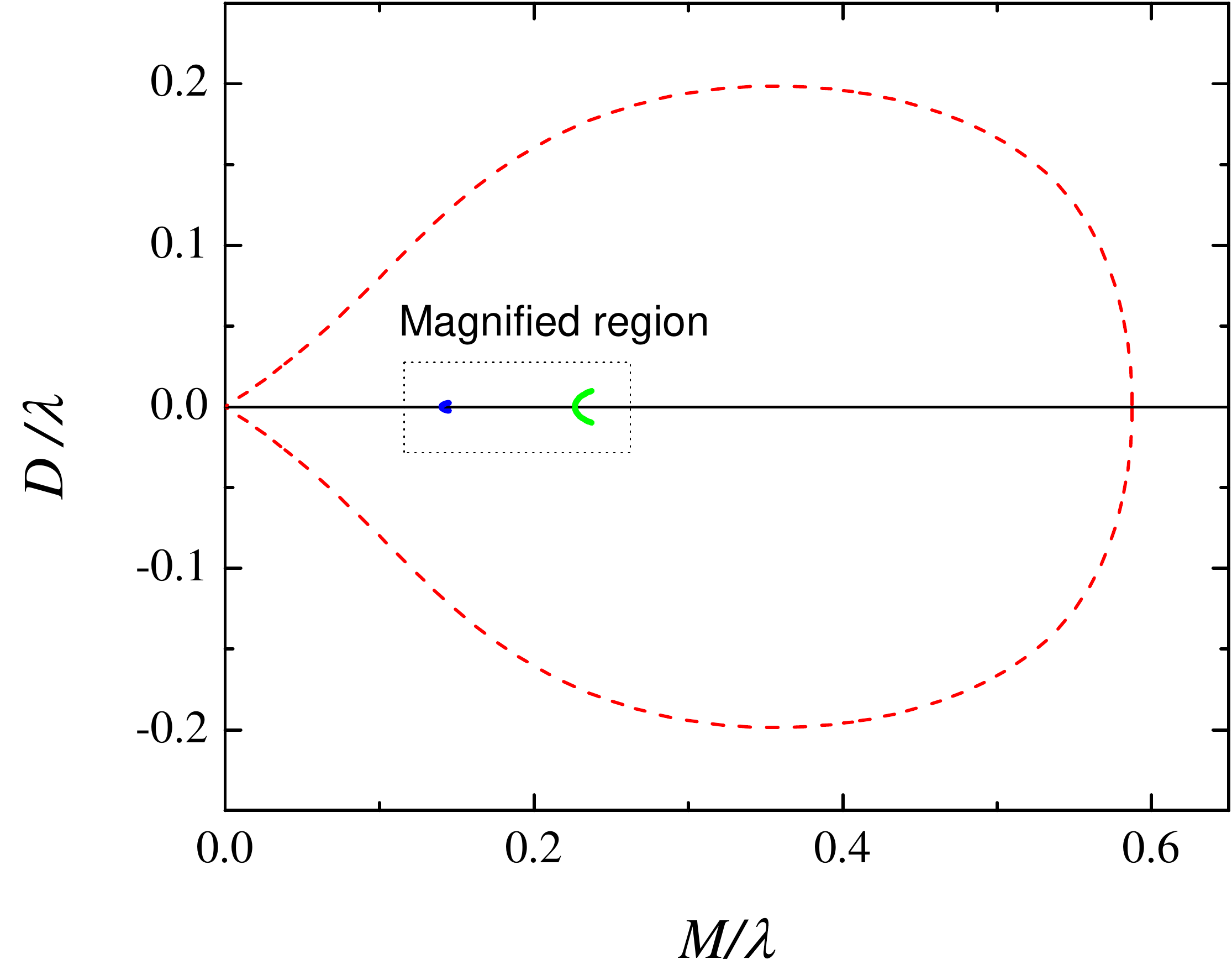}
	\includegraphics[width=0.495\textwidth]{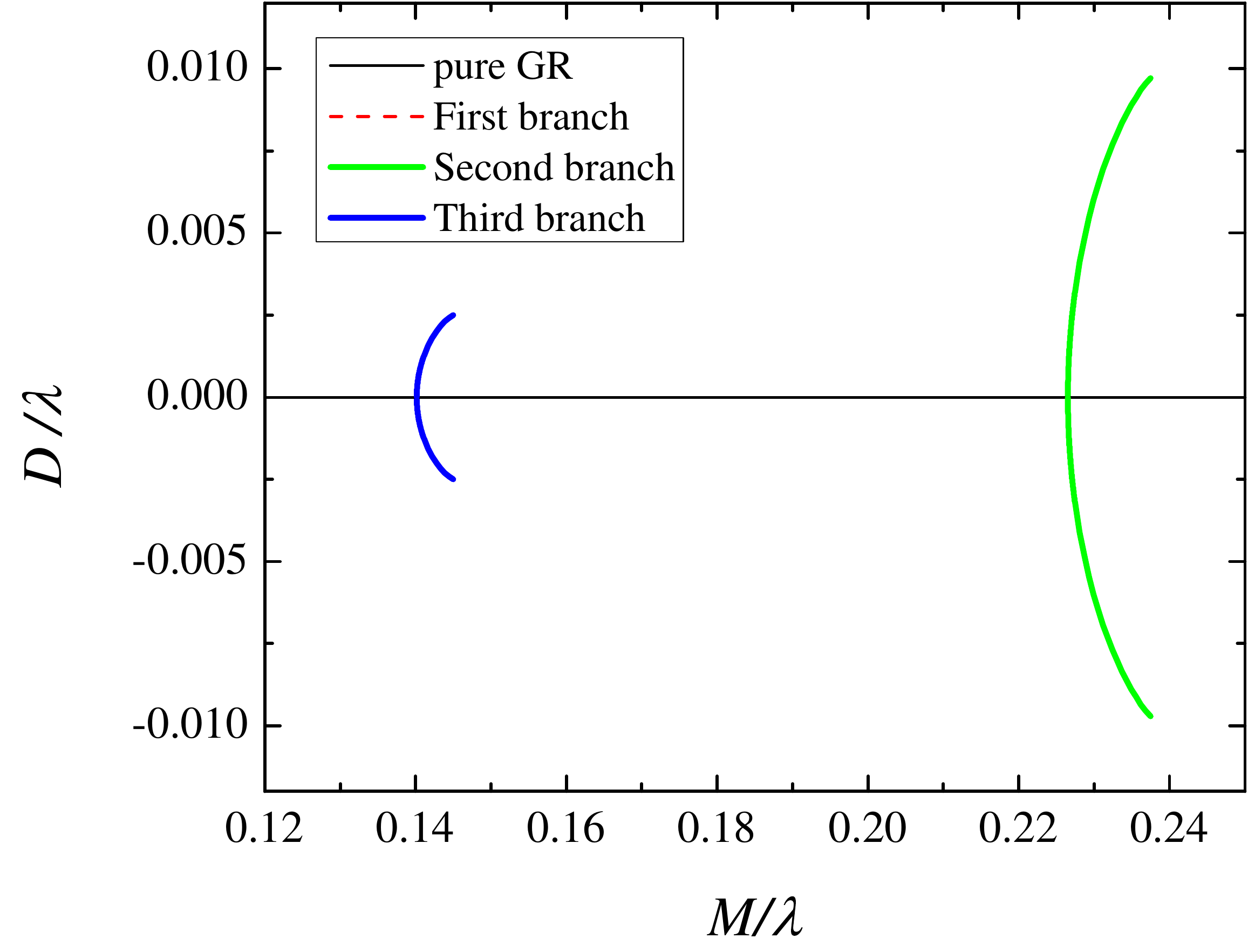}
	\caption{The dilaton charge of the black hole as a function of its mass. The right figure is a magnification of the left one.}
	\label{fig:D(M)}
\end{figure}

\begin{figure}[htb]
	\includegraphics[width=0.495\textwidth]{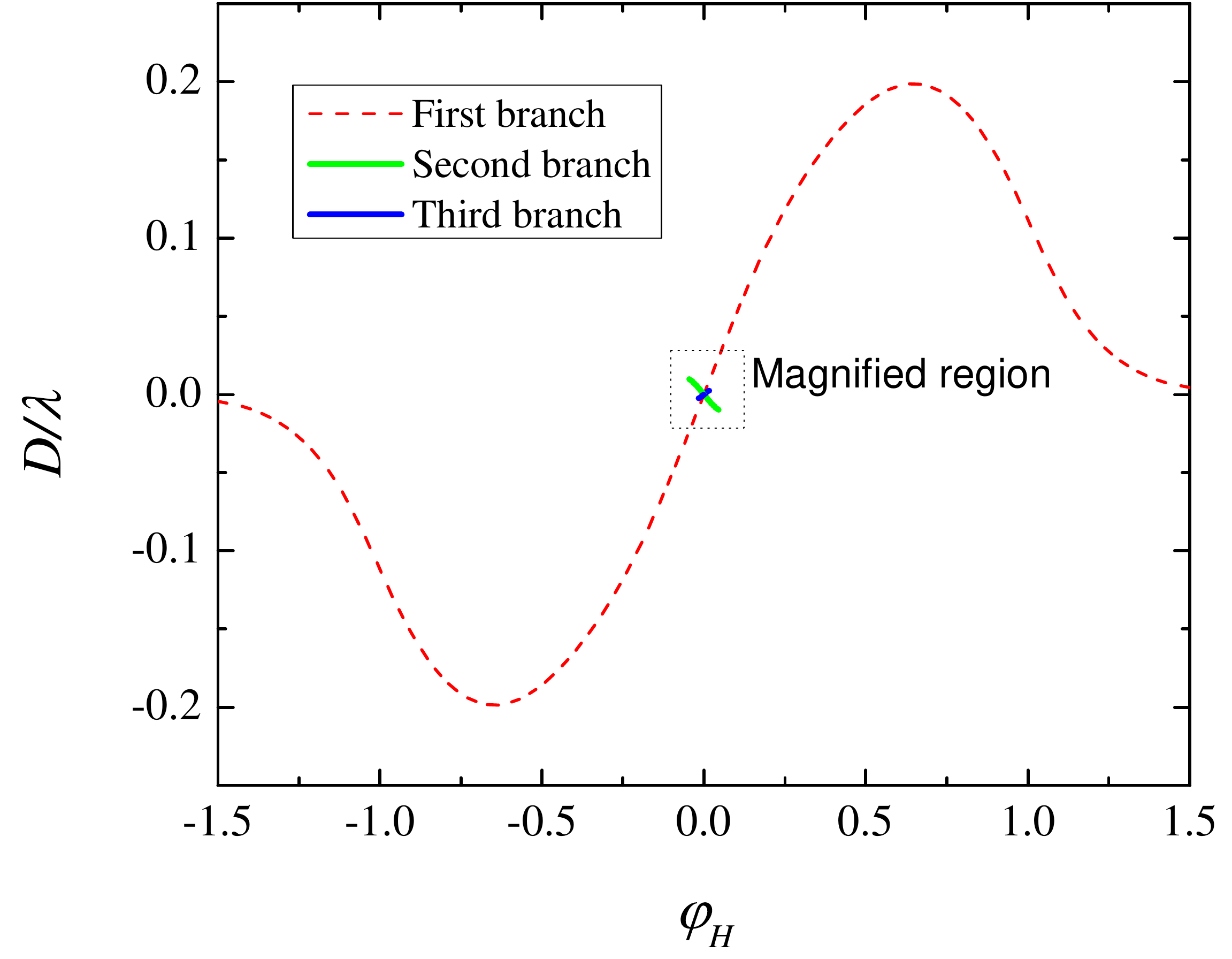}
	\includegraphics[width=0.495\textwidth]{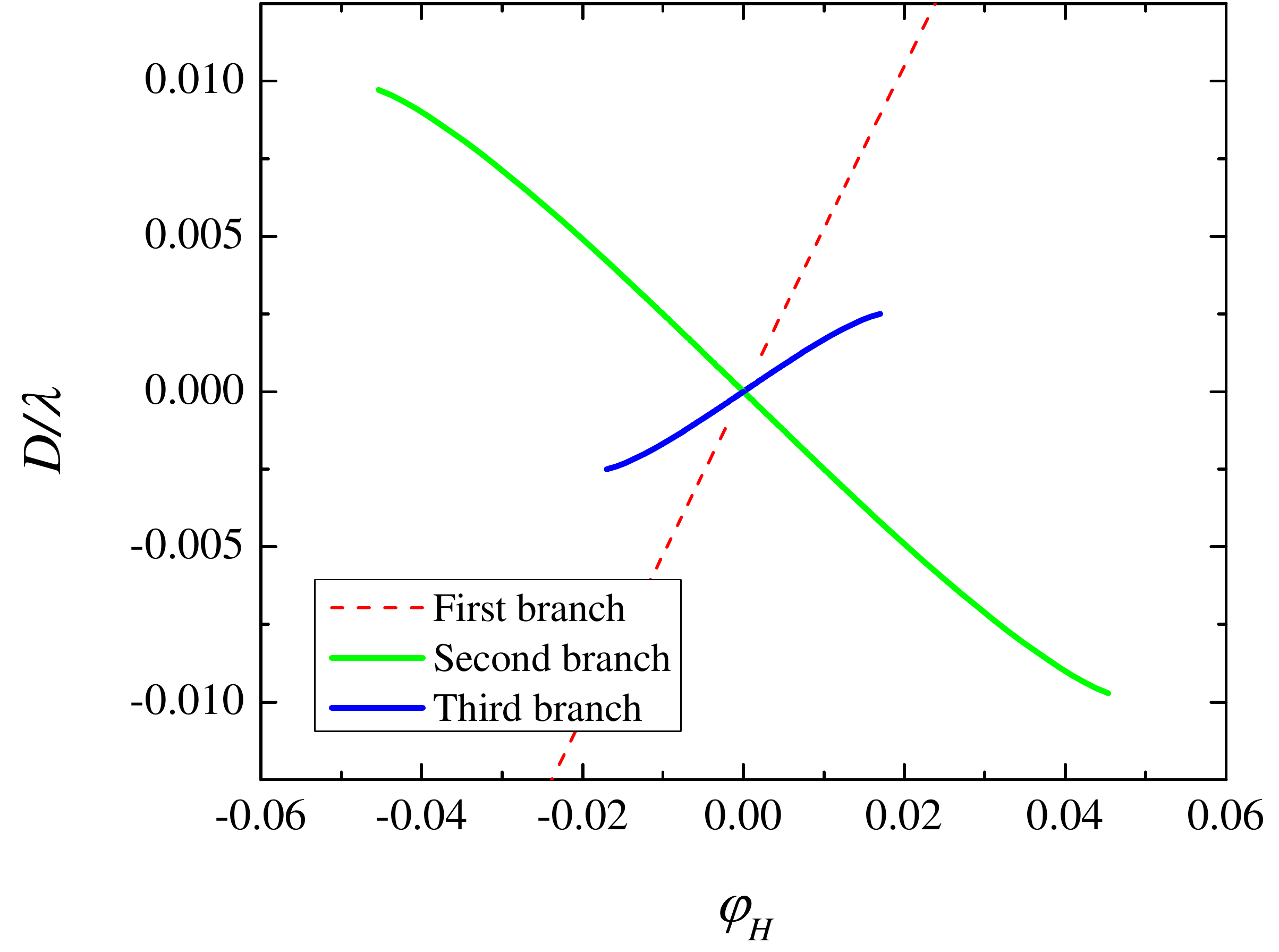}
	\caption{The dilaton charge of the black hole as a function of the scalar field at the horizon. The right figure is a magnification of the left one.}
	\label{fig:D(rh)}
\end{figure}

The dilaton charge as a function of the mass is shown in Fig. \ref{fig:D(M)} and as a function of $\varphi_H$ -- in Fig. \ref{fig:D(rh)}. As one can see, while the dependence $\varphi_H(M/\lambda)$ is monotonic for the first nontrivial branch and $\varphi_H$ increases significantly for small masses, $D(M/\lambda)$ has an extremum (either minimum or maximum depending on the sign of $\varphi_H$) and tends to zero for small masses.

\begin{figure}[htb]
	\includegraphics[width=0.495\textwidth]{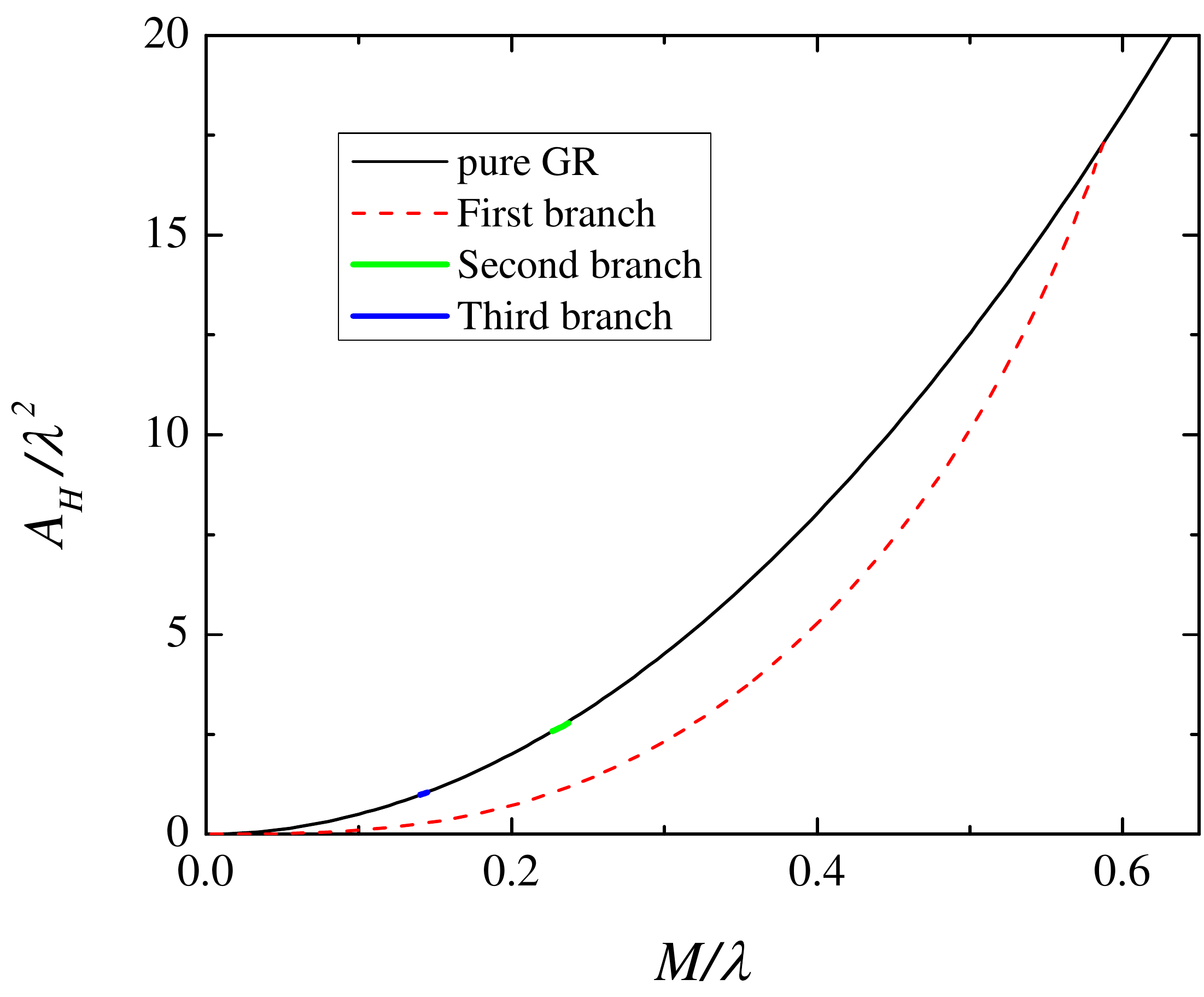}
	\includegraphics[width=0.495\textwidth]{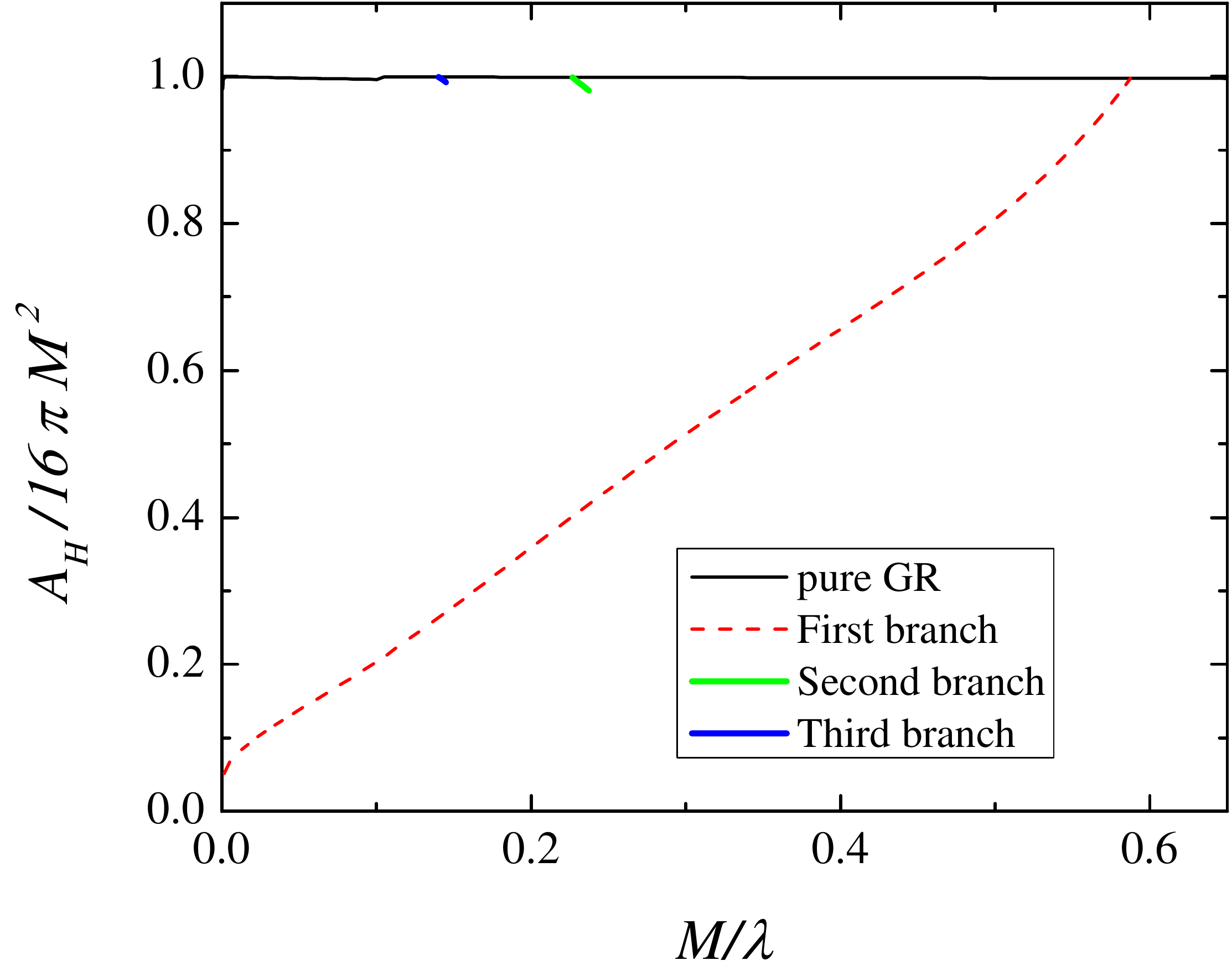}
	\caption{The area of the black hole horizon $A_H$ as a function of the mass. In the right figure the black hole area is normalized to the corresponding value in the Schwarzschild limit, i.e. $A_{H}^{Schwarschild}=16\pi M^2$.}
	\label{fig:Ah(M)}
\end{figure}

The area of the black hole horizon, $A_H=4\pi r_H^2$ and the normalized $A_H/A_{H}^{Schwarschild}$, where $A_{H}^{Schwarzschild}$ is the corresponding area of the Schwarzschild black hole  $A_{H}^{Schwarzschild}=16\pi M^2$, are plotted as functions of the mass in Fig. \ref{fig:Ah(M)}. The first three nontrivial branches of black holes are plotted in addition to the corresponding dependence in the Schwarzschild case. This graph can be used to better judge how strong the deviations from pure general relativity are. As one can see, only the first branch of nontrivial solutions (with scalar field which has no zeros) deviates significantly from the Schwarzschild case and the deviations are the largest for intermediate masses. This observation is similar to the behavior of the dilaton change which tends to zero for very large and very small masses, reaching maximum for intermediate values of $M/\lambda$. 

\begin{figure}[htb]
	\includegraphics[width=0.6\textwidth]{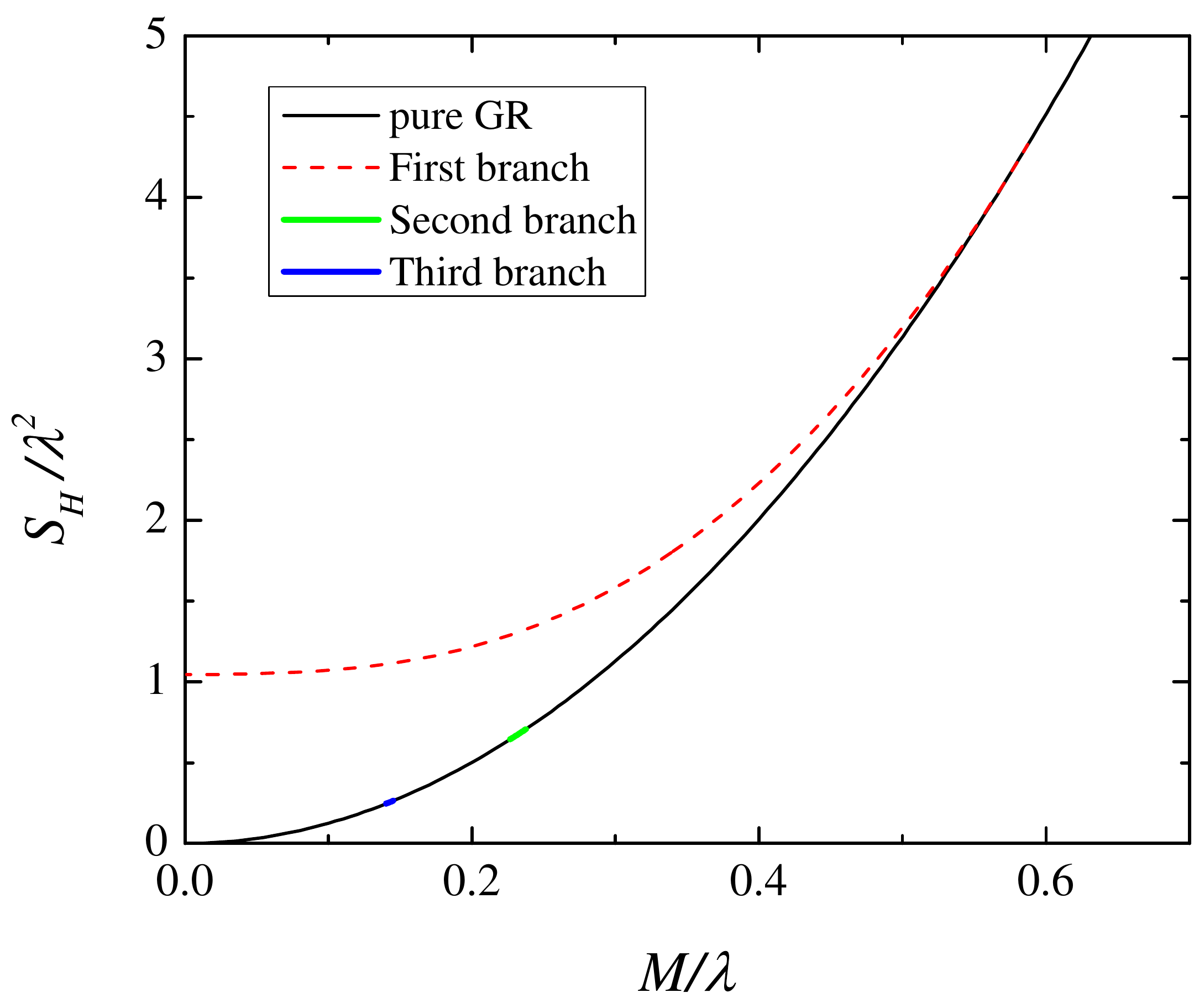}
	\caption{The entropy of the black hole as a function of its mass. .}
	\label{fig:SH(M)}
\end{figure}

Up to now we have shown and discussed the first three branches of nontrivial solutions and as we commented there are more branches that bifurcate at smaller values of $M/\lambda$. In order to have an indicator  for the stability of the black hole branches one can study the entropy of the black holes. The black hole entropy in the presence of a Gauss-Bonnet term in the action (\ref{eq:quadratic}) is not just one forth of the horizon area and its definition is a little bit more complicated. We adopt the entropy formula proposed by Wald in \cite{Wald_1993},\cite{Iyer_1994}  namely 

\begin{eqnarray}
S_{H}= 2\pi\int_{H} \frac{\partial {\cal L}}{\partial R_{\mu\nu\alpha\beta}} \epsilon_{\mu\nu}\epsilon_{\alpha\beta} ,
\end{eqnarray}
where ${\cal L}$ is the Lagrangian density and  $\epsilon_{\alpha\beta}$ is the volume form on the 2-dimensional cross section $H$ of the horizon.  In our case we find 
 \begin{equation}
S_H = \frac{1}{4} A_H + 4\pi \lambda^2 f(\varphi_{H}).
\end{equation}
The entropy as a function of the black holes mass is plotted in Fig. \ref{fig:SH(M)}. As one can see the first nontrivial branch has  entropy larger than the Schwarschild one and it is therefore thermodynamically more stable. This is an expected results since for masses smaller that the point of the first bifurcation the Schwarschild solution get unstable and there should be another one. The second and the third nontrivial branches on the other hand have lower entropy compared to the pure general relativistic case, which means that they are most probably unstable. The same is expected to apply for the rest of nontrivial branches that exist for smaller masses. The dynamical stability of our black hole solutions will be investigated in a future work.

\section{Conclusion}
In the present paper we have studied  black hole solutions in a particular class of ESTGB theories described by a coupling function $f(\varphi)$ that satisfies the conditions $\frac{df}{d\varphi}(0)=0$ and $\frac{d^2f}{d\varphi^2}(0)>0$. We have shown that for such theories an effect similar to the spontaneous scalarization of neutron stars exists -- the Schwarzschild solution becomes unstable below certain mass and new branches of black hole solutions with nontrivial scalar field appear that bifurcate from the Schwarzschild one at certain masses. The first branch of nontrivial solutions is characterized by a scalar field that has no zeroes while the scalar field has one zero for the second branch, two zeros for the third branch and so on. The general expectation, though, is that only the first branch of solutions would be stable and the rest would be unstable.  The main difference with the spontaneous scalarization of neutron stars is that the scalar field is not sourced the by matter, but instead by the extreme curvature of the spacetime around black holes. This places the considered solutions amongst the very few examples of scalarized black holes. 

We have explicitly constructed such solutions with nonzero scalar field and it was shown that the first branch of nontrivial solutions is thermodynamically more stable compared to the Schwarzshild one. The results presented in the current paper are for a particular coupling function that can produce non-negligible deviations from pure general relativity. We have tested, though, several other functions satisfying the above given conditions for $f(\varphi)$ and the results are qualitatively very similar.

It was demonstrated that the behavior of the scalar field at the horizon $\varphi_H$ and the dilaton charge $D$ (i.e. the coefficient in front of the $1/r$ term in the asymptotic expansion of the scalar field at infinity) are qualitatively different. While $\varphi_H$ monotonically increase with the decrease of the mass (for positive $\varphi_H$), the dilaton charge first increases and after reaching a maximum it decreases to zero. As expected, the $A_H(M)$ dependence exhibits similar behavior to $D$, i.e. the black hole solutions tend to the Schwarzschild one for very small masses and for larger masses close to the bifurcation point and the maximum deviation is observed for intermediate masses. We should note that the above given observations are true only for the first nontrivial branch characterized by scalar field without zeros. The rest of the branches are terminated at some nonzero mass because beyond that mass they violate  condition (\ref{eq:BC_sqrt_rh}). 

We have studied the behavior of the black hole entropy and the results show that the first nontrivial branch has higher entropy than the Schwarzschild black holes and it is thermodynamically more stable while the rest of the branches have lower entropy. Thus, the general expectation is that the first branch of solutions is stable and it is the one that would realize in practice because of the instability of the Schwarzschild solutions. The other nontrivial branches are supposed to be unstable. Of course, this can be rigorously proven only if one considers the linear perturbations of the solutions with nontrivial scalar field that would be done in a future publication.  

Finally,  let us briefly comment on the following. The black holes we are considering posses a nontrivial scalar field, and thus they have scalar ``hair''. When the branch of the solution is fixed then this  ``hair'' is secondary which  means that the dilaton charge is not an independent parameter but instead it depends on black hole mass. However, the number of the branches is an independent 
parameter introducing a new ``hair''  of discrete type.  One  may adopt the view that  only the stable branch  has to be considered  
getting rid in this way of the discrete  ``hair''. In our opinion the classification of black hole solutions presented in the present work
is rather subtle and needs a much deeper analytical investigation. We shall discuss this problem in a future publication.

\section*{Acknowledgements}
DD would like to thank the European Social Fund, the Ministry of Science, Research and the Arts Baden-W\"urttemberg for the support. DD is indebted to the Baden-Württemberg Stiftung for the financial support of this research project by the Eliteprogramme for Postdocs. The support by the Bulgarian NSF Grant DFNI T02/6, Sofia University Research Fund under Grants 80.10-30/2017 and 3258/2017, and 
COST Actions MP1304, CA15117, CA16104   is also gratefully acknowledged.

%%%%%%%%%%%%%%%%%%%%%%%%%%%%%%%%%%%%%%%%%%%%%%%%%%%%%%%%%%%%%%%%%%%%%%%%%%%%%%%

\end{document}